\documentclass[10pt,letterpaper]{article}
\usepackage{opex3}
\usepackage{amsmath}
\usepackage{cite}

\def\XXint#1#2#3{{\setbox0=\hbox{$#1{#2#3}{\int}$}
\vcenter{\hbox{$#2#3$}}\kern-.5\wd0}}

\begin{document}
\title{Flattening axial intensity oscillations of a diffracted Bessel beam through a cardioid-like hole }

\author{Jia-Sheng Ye,$*$ Li-Juan Xie, Xin-Ke Wang, Sheng-Fei Feng, Wen-Feng
Sun, and Yan Zhang}
\address{Department of Physics, Capital Normal University,\\
Beijing Key Laboratory of Metamaterials and Devices,\\
Beijing Advanced Innovation Center for Imaging Technology,\\
Key Laboratory of Terahertz Optoelectronics, Ministry of Education, Beijing 100048, P. R.
China}
\email{$^*$Corresponding author: jsye@mail.cnu.edu.cn}

\date{\today}

\begin{abstract}
We present a new feasible way to flatten the axial intensity
oscillations for diffraction of a finite-sized Bessel beam, through
designing a cardioid-like hole. The boundary formula of the cardioid-like
hole is given analytically. Numerical results by the complete Rayleigh-Sommerfeld
method reveal that the Bessel beam propagates
stably in a considerably long axial range, after passing through the
cardioid-like hole. Compared with the gradually absorbing apodization
technique in previous papers, in this paper a hard truncation of the
incident Bessel beam is employed at the cardioid-like hole edges. The
proposed hard apodization technique takes two advantages in
suppressing the axial intensity oscillations, i.e., easier
implementation and higher accuracy. It is expected to have practical
applications in laser machining, light sectioning, or optical
trapping.
\end{abstract}

\section{Introduction}
The nondiffracting beam exhibits a strict propagating invariance
property in free space with good axial intensity uniformity as
well as stable transverse resolution \cite{Bouchal:2003}. Therefore,
it attracts intensive research interest due to its wide applications
in laser machining \cite{Rioux:AO1978}, light sectioning
\cite{Hausler:AO1988}, interferometry \cite{Brinkmann:Optik1996},
optical trapping \cite{Ambriz:OE2013,Cizmar:NP2010}, microscopy \cite{Fahrbach:NP2010} and so on. Durnin {\sl et al.}
have demonstrated mathematically that the three-dimensional
infinite-sized Bessel beams are exact solutions to the scalar wave
equations \cite{Durnin:PRL1987,Durnin:JOSAA1987}. However, it is
physically impossible to generate an infinite-sized Bessel beam, in
that it is not square integrable and contains infinite energy.
Approximate Bessel beams were created in experiments by using an
annular slit in the back focal plane of a lens
\cite{McQueen:AJP1999,Mohamed:AO2013,Cizmar:OE2009}, holographic techniques
\cite{Cizmar:OE2009,Turunen:AO1988,Vasara:JOSAA1989,Tao:AO2004}, axicons
\cite{Popov:JOA1998,Popov:OC1998}, or other ways
\cite{Lapointe:JOLT1992}. In many previous papers, it was reported
that the Bessel beam diffracted from a finite-sized circular hole
presented prominent axial intensity oscillations
\cite{Herman:AO1992,Cox:OL1992,Jiang:AO1995}. A sudden
cut-off of the incident field leads to the strong diffraction effect
at the hole edges. This problem was generally solved by
covering a gradually absorbing mask. Almost complete suppression of
axial intensity oscillations was achieved through apodization of the
incident Bessel beam with flattened Gaussian or trigonometric
apodized functions
\cite{Cox:OL1992,Borghi:JOSAA1997,Jaroszewicz:OL1993}.

Although the gradually absorbing apodization technology presents a flat
axial intensity profile theoretically, it faces two problems in
practical applications. Firstly, the amplitude modulation is
usually realized by using the spatial light modulator (SLM)
\cite{Goodman:1996,DelaTocnaye:AO1997,Song:USpatent2013,Davis:OL1999,Arrizon:OE2005},
but the transmittance is difficult to be adjusted exactly the same
as the expected value. If each pixel on the SLM brings about a
transmittance error, the obtained intensity distribution will deviate
much away from the ideal one. Secondly, it would be difficult that
both the total size and spatial resolution of the SLM meet
requirements simultaneously so that the apodized field drops very
smoothly around the aperture edges. Based on these two
considerations, a question is raised. Can we flatten the
axial intensity oscillation of a finite-sized Bessel beam by a hard apodization
(HA) technique? Here the HA technique
means that there only exists a transmitted hole on the input plane,
with other parts being opaque. Gray-level transmittance
is no longer used for relieving the experimental difficulty and
heightening the achieved intensity accuracy. For comparison, we
refer to the gradually absorbing apodization technology as the soft
apodization (SA) technique.

This paper is organized as follows. In Section 2, the design
principles are described in detail, and the boundary formula of the
transmitted hole is deduced analytically. For demonstrating the
validity of our proposed strategy, in Section 3 the parameters are
given and wave propagating behaviors are simulated by the complete
Rayleigh-Sommerfeld method \cite{Gillen:2004}. Physical
explanations are also delivered. Section 3 consists of two
subsections. In subsection 3.1, axial intensity distributions are
calculated, and the intensity uniformity is characterized.
In subsection 3.2, intensities on several cross-sectional planes
within the beam propagation range are
evaluated. A brief conclusion is
drawn in Section 4 with some discussions.

\section{Boundary formula of the transmitted hole for hard apodization}
\label{sec:theory} Let's first consider the diffraction of a Bessel
beam from a finite-sized circular hole. The input plane is the $xy$
plane, which situates at $z_0=0$ m. The incident Bessel beam
propagates along the $z$-axis, whose amplitude distribution
satisfies the $J_0(\beta r)$ function, where $J_0$ is the zeroth
order Bessel function; $\beta$ is the propagation constant;
$r=\sqrt{x^2+y^2}$ denotes the radial distance from an arbitrary source point
$(x,y)$ to the origin on the input plane.

In the transmitted region ($z>0$), the field at any point
$(x',y',z)$ is calculated by the complete Rayleigh-Sommerfeld method as \cite{Gillen:2004}
\begin{equation}
E(x',y',z)=\frac{kz}{i2\pi}\int\int_{\Omega}
E_0(x,y,0)\frac{\exp(ik\rho)}{\rho^2}(1-\frac{1}{ik\rho})\mathrm{d}x\mathrm{d}y\,,
\label{eq:integral}
\end{equation}
where $k=2\pi/\lambda$ represents the wave vector and $\lambda$ represents the
incident wavelength; $\Omega$ denotes integral region of the circular hole;
$\rho=\sqrt{(x'-x)^2+(y'-y)^2+z^2}$ is the distance from the source point $(x,y,0)$ on
the input plane to the observation point $(x',y',z)$.
$E_0(x,y,0)=J_0(\beta r)=J_0(\beta\sqrt{x^2+y^2})$ stands for the incident field. The intensity at the
observation point is given by $I(x',y',z)=|E(x',y',z)|^2$, where $|E|$ represents magnitude of
the complex number $E$. On assuming
$x'=y'=0$, we will obtain the axial intensity distribution of the
Bessel beam diffracted from the circular hole. In this case, the axial field can be simplified in
polar coordinates as follows:
\begin{equation}
E(0,0,z)=(-ikz)\int_{0}^{R}E_0(r,0)\frac{\exp(ik\sqrt{r^2+z^2})}{r^2+z^2}[1-\frac{1}{ik\sqrt{r^2+z^2}}]r\mathrm{d}r\,,
\label{eq:polarfield}
\end{equation}
where $R$ is the radius of the circular hole; $E_0(r,0)=J_0(\beta r)$ represents the incident Bessel beam on the input plane z=0;
$r=\sqrt{x^2+y^2}$ denotes the radial distance on the input plane.

\begin{figure}[htb]
\centering\includegraphics[width=13.2cm]{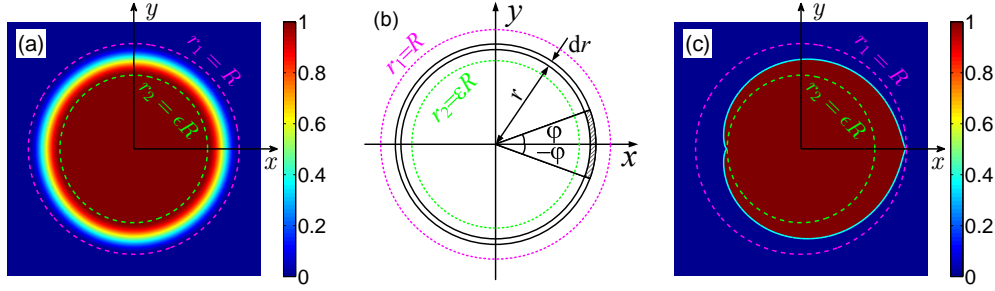} \caption{(a) The
transmittance distribution on the input plane $z_0=0$ m under the
soft apodization (SA) condition. The green and pink dashed curves
depict two circles with radii of $\epsilon R$ and $R$,
respectively. (b) The shadowed annular district represents the
equivalent transmitted region at a definite radius $r$, which
occupies 2$|\varphi|$ in the angle direction. (c) The transmittance
distribution on the input plane under the hard apodization (HA)
condition. The cyan solid curve encloses a cardioid-like hole, with
transmittance being 1 and 0 inside and outside the hole,
respectively. }\label{fig:shematic}
\end{figure}

It is well known that strong axial intensity oscillations are encountered due to a hard
truncation of the incident field. For suppressing the axial
intensity oscillations, in Ref. \cite{Cox:OL1992} Cox {\sl et al.}
introduced a trigonometric apodized function as follows
\begin{eqnarray}
T(r)= \begin{cases} 1, & r<\epsilon R,\cr
\displaystyle{\frac{1+\cos[\pi(r-\epsilon R)/(R-\epsilon R)]}{2}}, &
\epsilon R\leq r\leq R, \cr 0, & r>R,
\end{cases}
\label{eq:transmittance}
\end{eqnarray}
where $T(r)$ represents the amplitude transmittance function on the input
plane; $\epsilon$ is a smoothing parameter ranging from 0 to 1.
Under this SA condition, the transmittance gradually decreases from
1 to 0 within an annular region $r\in[\epsilon R, R]$, as displayed
in Fig. \ref{fig:shematic}(a). The green and pink dashed curves
depict two circles with radii of $\epsilon R$ and $R$,
respectively. It is seen from Fig. \ref{fig:shematic}(a) that the
color gradually becomes from red to blue when the radius increases
from $\epsilon R$ to $R$. Correspondingly, the transmittance
monotonically decreases from 1 to 0. For calculating the transmitted axial 
intensity, Eq. (2) is used where the input field now becomes $E_0(r,0)=J_0(\beta r)T(r)$. 
By using this SA technique,
they achieved a flat axial intensity distribution of the
diffracted Bessel beam, as seen in Figs. 4(a) and 4(b) in Ref.
\cite{Cox:OL1992}.

In the following, we will discuss our strategy. From Eq.
(\ref{eq:polarfield}), the field at any axial point $(0,0,z)$ is a
diffraction superposition of all the source points on the input
plane. Since there is only one integral variable $r$ in Eq. (\ref{eq:polarfield}),
the optical system is rotationally symmetric. Namely, in the input plane
each source point on a definite circle
with the same radius $r$ has the same incident field $E_0(r,0)=J_0(\beta r)$,
contributing equally to a given axial observation point. Therefore,
we can make such a equivalent change. For instance, in Fig. 1(a), if
the transmittance $T(r_0)$ equals 0.5 for some radius $r_0$, we may
let the incident field passes one half while the other half is
blocked at radius $r_0$. It is easy to understand that we should get the
same axial intensity distributions in both cases. For other
transmittance values, the only difference lies in the filling ratio
of the transmitted part.

Now we turn to explore the general shape that the transmitted hole
should be like. In Fig. \ref{fig:shematic}(b), the shadowed region
represents the equivalent transmitted part for a given radius $r$
within $r\in[\epsilon R, R]$. The filling ratio $F(r)$ should fulfill
the following equation
\begin{equation}
\displaystyle{F(r)=\frac{2|\varphi|}{2\pi}}=T(r)=\displaystyle{\frac{1+\cos[\pi(r-\epsilon
R)/(R-\epsilon R)]}{2}},  \hskip1cm
-\pi\leq\varphi\leq\pi.\label{eq:fillingratio}
\end{equation}
After mathematical transformation, we may write Eq.
(\ref{eq:fillingratio}) explicitly as follows
\begin{equation}
r(\varphi)=\epsilon R+\displaystyle{\frac{(R-\epsilon
R)\arccos(2|\varphi|/\pi-1)}{\pi}}, \hskip1cm
-\pi\leq\varphi\leq\pi.\label{eq:curve}
\end{equation}
All the points satisfying Eq. (\ref{eq:curve}) form a closed curve,
as drawn by the cardioid-like cyan solid curve in Fig.
\ref{fig:shematic}(c). From the above analysis, if the incident Bessel beam
passes through a cardioid-like hole while it is prevented
outside, we can expect the same flat axial intensity distribution
as that in Ref. \cite{Cox:OL1992}. Figure
\ref{fig:shematic}(c) shows the transmittance distribution on the
input plane under the proposed HA condition, indicating the
transmittance being 1 and 0 inside and outside the region encircled
by the cardioid-like curve, respectively.

\section{Propagating behaviors of the apodized Bessel beams}
\label{sec:results} In this section, we will perform numerical
simulations to prove the validity of the proposed HA technique.
Parameters are chosen as follows. For the incident Bessel beam, the
propagation constant is $\beta=10^4~\mathrm{m}^{-1}$ with wavelength
of $\lambda=0.5~\mu\mathrm{m}$. The circular hole has a radius of
$R=50~\mathrm{mm}$.

\subsection{The axial intensity distribution of the apodized Bessel beam}
\begin{figure}[htb]
\centering\includegraphics[width=13.2cm]{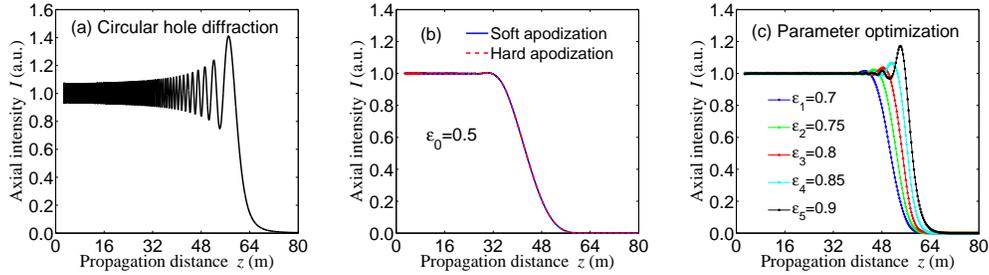} \caption{(a) The
axial intensity distribution of the Bessel beam diffracted from a
circular hole without apodization. (b) The blue solid and red dashed
curves represent the axial intensity distributions of the diffracted
Bessel beam by the SA and HA techniques, corresponding to Figs.
\ref{fig:shematic}(a) and \ref{fig:shematic}(c), respectively. In
both cases, we choose the same smoothing parameter as
$\epsilon_0=0.5$. (c) Axial intensity distributions of the
diffracted Bessel beam by the HA technique, when the smoothing
parameter $\epsilon$ varies from 0.7 to 0.9 with an interval of
0.05. The blue, green, red, cyan, and black curves correspond to
different smoothing parameters of 0.7, 0.75, 0.8, 0.85, and 0.9,
respectively. }\label{fig:axial}
\end{figure}

In this subsection, the axial intensity distribution of the
diffracted Bessel beam is investigated. Firstly, by assuming
$E_0(r,0)$ to be $J_0(\beta r)$ in Eq. (\ref{eq:polarfield}), we can
obtain the axial intensity distribution of the diffracted Bessel
beam from a finite-sized circular hole without apodization, as shown
by the black curve in Fig. \ref{fig:axial}(a). It is seen from Fig.
\ref{fig:axial}(a) that the axial intensity oscillates with a
relative error of about $\pm7\%$. Secondly, axial intensity profiles
are flattened by the SA and HA techniques to prove their
equivalence, as shown by the blue solid and red dashed curves in
Fig. \ref{fig:axial}(b), respectively. In both cases, we choose the
same smoothing parameter $\epsilon_0$ as 0.5. Under the SA
condition, the incident field on the input plane in Eq.
(\ref{eq:polarfield}) is given by $E_0^\mathrm{s}(r,0)=J_0(\beta
r)\times T(r)$, where $T(r)$ is given by Eq.
(\ref{eq:transmittance}). The integral region $\Omega_1$ is the circular 
hole. Under the HA condition, since the cardioid-like hole is
rotationally asymmetric, axial intensity should be calculated from Eq.
(\ref{eq:integral}), where the incident field on the input plane is
$E_0(x,y,0)=J_0(\beta\sqrt{x^2+y^2})$ and the integral region $\Omega_2$ is
the cardioid-like hole determined by Eq. (\ref{eq:curve}). It is
demonstrated in Fig. \ref{fig:axial}(b) that both curves overlap,
namely, we have achieved the same flat axial intensity
distributions of the diffracted Bessel beam by the
previous SA and the proposed HA techniques.
Thirdly, through optimizing the smoothing parameter $\epsilon$ in
Eq. (\ref{eq:curve}), we have computed the diffracted axial
intensity distributions of the Bessel beam enclosed by the cardioid-like curve,
as drawn in Fig. \ref{fig:axial}(c). The smoothing parameter
$\epsilon$ ranges from 0.7 to 0.9 for every 0.05. The blue, green,
red, cyan, and black curves correspond to different smoothing
parameters of 0.7, 0.75, 0.8 ,0.85, and 0.9, respectively. With the
increase of the smoothing parameter, the propagation distance is
lengthened while the axial intensity uniformity is weakened, as
observed in Fig. \ref{fig:axial}(c). It can be interpreted. When the
smoothing parameter becomes larger, the transmittance drops more
rapidly from 1 to 0 near the hole edges. Therefore, the diffraction
effect is stronger and the axial intensity oscillations appear to be
more prominent. When the smoothing parameter $\epsilon$ reaches 1,
it degenerates to the case in Fig. \ref{fig:axial}(a). On taking
both the axial intensity uniformity and the propagation distance
into account, the smoothing parameter is compromised as 0.8 in the
following.

Next, the axial intensity uniformity is quantitatively characterized when
the smoothing parameter is 0.8. The flat axial intensity region is defined as
the axial domain with intensity relative error less than $\pm0.01\%$. Numerical results reveal
that the flat axial intensity region extends out to $z_f=32.01$ m. The axial range expands
to 39.15 and 46.42 m when the restriction of intensity relative error is relaxed to $\pm0.1\%$ and $\pm1\%$, respectively.
Then, the axial intensity oscillations are gradually strengthened. After the axial intensity reaches its peak value of 1.037 at $z_f=48.44$m,
it drops rapidly towards 0. When the propagation distance is longer than 61.46 m, the axial intensity drops below 1\%.
The numerical simulations agree well with the results obtained from geometrical optics. Since the Bessel beam can be considered
as superposition of plane waves whose wave vectors lie on a cone with an angle $\theta$ with respect to the $z$-axis \cite{McQueen:AJP1999,Cox:OL1992}, where
$\beta=k\sin(\theta)=2\pi\sin(\theta)/\lambda$. In geometrical optics,
the effective axial beam range is given by
$z_{\mathrm{max}}=R/\tan(\theta)=62.83$ m.

\subsection{Intensity distributions on several cross-sectional planes for the hard apodized Bessel beam }

\begin{figure}[htb]
\centering\includegraphics[width=12cm]{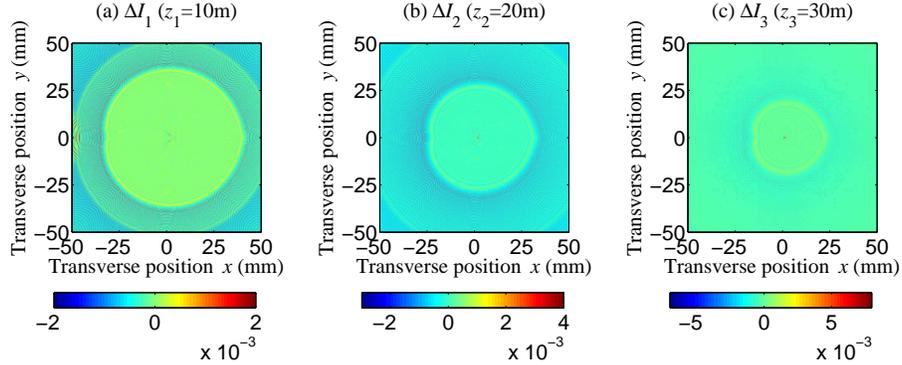} \caption{Intensity deviation on three cross-sectional planes
for the hard apodized Bessel beam. The smoothing parameter $\epsilon$ is 0.8, and other parameters
are the same as above. (a) The intensity deviation $\Delta I_1=I_1(x_1,y_1)-I_0(x,y)$, where
$I_1(x_1,y_1)$ represents the intensity distribution on the lateral
plane $z_1=10$ m and $I_0(x,y)=|J_0(\beta\sqrt{x^2+y^2})|^2$ is 
the intensity of the incident Bessel beam on the input
plane $z_0=0$ m. (b) and (c) are the same as (a) except for
$z_2=20$ m and $z_3$=30 m, respectively.}\label{fig:regional}
\end{figure}

Since the cardioid-like curve in Fig. \ref{fig:shematic}(c) is rotationally asymmetric,
the field on cross-sectional planes should also be asymmetric. In this case, the transmitted 
intensity distributions are calculated from Eq. (\ref{eq:integral}) and the integral region
$\Omega$ is the cardioid-like hole. To see how asymmetric the field
is, figure \ref{fig:regional} displays the regional intensity deviation on three
cross-sectional planes. Figure \ref{fig:regional}(a) presents the
intensity deviation $\Delta I_1=I_1(x_1,y_1)-I_0(x,y)$, where
$I_1(x_1,y_1)$ represents the intensity distribution on the lateral
plane $z_1=10$ m and $I_0(x,y)=|J_0(\beta\sqrt{x^2+y^2})|^2$ is 
the intensity of the incident Bessel beam on the input
plane $z_0=0$ m. Figures \ref{fig:regional}(b) and
\ref{fig:regional}(c) are the same as Fig. \ref{fig:regional}(a) except for
$z_2=20$ m, and $z_3=30$ m, respectively. A cardioid-like shape of intensity deviation is
observed in Fig. \ref{fig:regional}, which suggests that asymmetry of the cardioid-like hole
leads to the intensity asymmetry. With the propagation of Bessel beam, the size of the cardioid-like shape
is gradually decreased. The intensity deviation is in the magnitude of $10^{-3}$ from the color bars,
indicating the intensity relative error is less than 1\%.
Accordingly, the diffracted Bessel beam preserves a relatively stable propagation property in a considerably long
axial range, although somewhat asymmetric.

\begin{figure}[htb]
\centering\includegraphics[width=12cm]{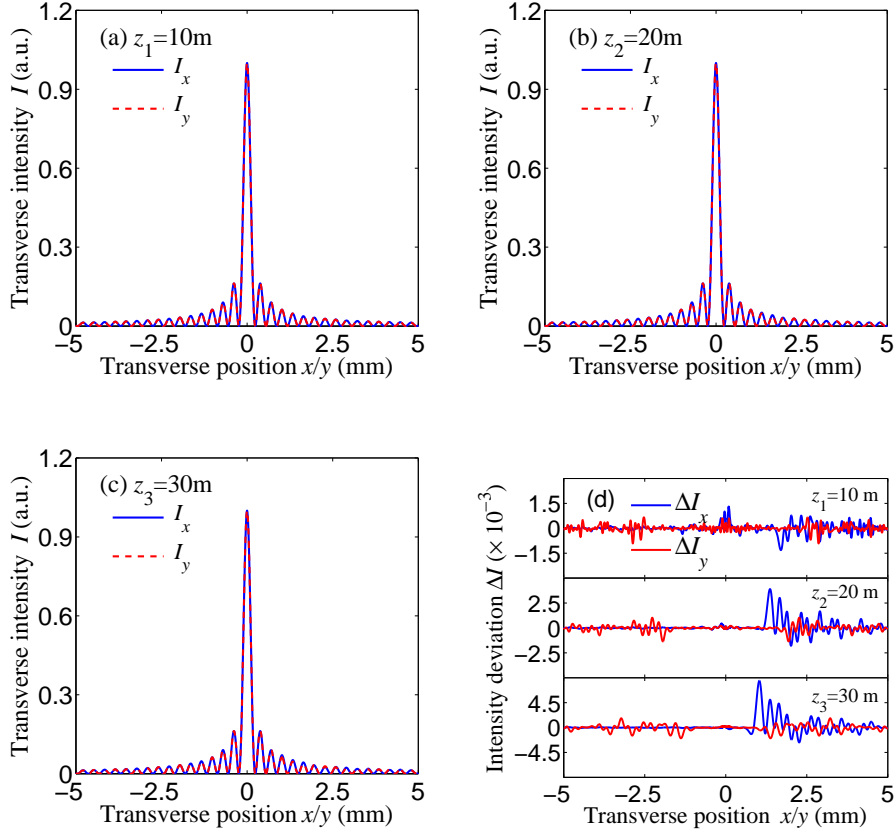} \caption{The
intensity distributions on three cross-sectional planes
with longitudinal coordinates of (a) $z_1$=10 m, (b) $z_2$=20 m, and
(c) $z_3$=30 m, respectively. The smoothing parameter $\epsilon$ is
assumed to be 0.8, and other parameters are the same as above. The
blue solid and red dashed curves represent the intensity profiles
along the $x$-axis and the $y$-axis, respectively. (d) Intensity
deviation $\Delta I_{x}=I_i(x_i,0)-|J_0(\beta |x|)|^2$ and $\Delta
I_y=I_i(0,y_i)-|J_0(\beta |y|)|^2$ on the above-mentioned three
cross-sectional planes are displayed from top to bottom, where $I_i(x_i,0)$ and
$I_i(0,y_i)$ (i=1,2,3) represent the intensities along the $x$-axis and the
$y$-axis, respectively. }\label{fig:transverseplot}
\end{figure}

To see the intensity asymmetry property more clearly, we extract
the intensity distributions along the $x$-axis and the $y$-axis
on the above three cross-sectional planes, as shown in Figs. \ref{fig:transverseplot}(a),
\ref{fig:transverseplot}(b), and \ref{fig:transverseplot}(c).
The blue solid and red dashed curves correspond to
intensities along the $x$-axis and the $y$-axis, respectively. It is
seen from Figs. \ref{fig:transverseplot}(a) to
\ref{fig:transverseplot}(c) that the intensity profiles on both axes
almost overlap. In order to magnify the intensity difference along the $x$-axis
and the $y$-axis, we calculate the intensity
deviation $\Delta I_{x}=I_i(x_i,0)-|J_0(\beta |x|)|^2$ and $\Delta
I_y=I_i(0,y_i)-|J_0(\beta |y|)|^2$ on the three
cross-sectional planes, as shown by the blue and red curves in Fig.
\ref{fig:transverseplot}(d) from top to bottom. $I_i(x_i,0)$ and
$I_i(0,y_i)$ (i=1,2,3) represent the intensities along the $x$-axis and the
$y$-axis, respectively. The intensity deviation $\Delta I_{x}$ or $\Delta I_{y}$
presents the difference between the intensity at the observation point and that at
the corresponding point on the input plane, indicating the propagation invariance property.
It is seen from Fig. \ref{fig:transverseplot}(d) that the red curves ($\Delta I_y$) are symmetric about
the origin point while the blue curves ($\Delta I_x$) are asymmetric about the origin point.
The reason is that the cardioid-like hole is symmetric about the $x$-axis but asymmetric about
the $y$-axis.  Since the maximum intensity deviation
is less than $\pm10^{-2}$, the diffracted Bessel beam propagates relatively
stably within the beam propagation range. If we
calculate the asymmetric intensity difference $\Delta I_{xy}=I_x-I_y=\Delta I_x-\Delta
I_y$, the intensity asymmetric property will be quantified. The intensity asymmetry does exist since
the two curves in Fig. \ref{fig:transverseplot}(d) are clearly separated. However, the
maximum asymmetric intensity difference $\Delta I_{xy}$ is only 0.89\%. Especially, for
applications of Bessel beams we are generally interested in a few lobes near the center.
The asymmetric intensity difference $\Delta I_{xy}$ is below 0.12\% for the central five intensity lobes.
Consequently, the intensity asymmetry can almost be neglected in the paraxial region.

\section{Conclusion and discussions}
\label{sec:conclusion} In this paper, we present a new feasible way
to flatten the diffracted axial intensity distribution of a
finite-sized Bessel beam. Numerical results by the complete
Rayleigh-Sommerfeld method have validated our strategy. It is found
that the Bessel beam diffracted from a cardioid-like hole
maintains a very good propagating invariance property along the axial
direction. The boundary formula of the cardioid-like hole is
given analytically. Compared with the previous SA technique, the
proposed HA technique provides a more feasible choice in practical
applications with higher accuracy, since the error only comes from
the hole boundary.

In fact, the proposed HA technique is a universal method for suppressing the axial
intensity oscillations, by transforming the transmittance
distribution into spatial transmitted filling factors. Therefore,
for other amplitude transmittance functions like the flattened Gaussian shape, by
analogy we can obtain another closed transmitted hole on the input
plane. Moreover, the proposed method may be extended to suppress
axial intensity oscillations for other kinds of incident waves, such
as plane waves or Gaussian waves. It is expected to have practical
applications in many optical processing systems.

\section*{Acknowledgements}{This work was supported by the 973 Program of China (No.
2013CBA01702); the National Natural Science Foundation of China (No.
11374216, 11404224, 11774243, 11774246, and 11474206); General Program of Science and Technology
Development Project of Beijing Municipal Education Commission under Grant No. KM201510028004;
Beijing Youth Top-Notch Talent Training Plan (CIT\&TCD201504080); Beijing Nova Program Grant No.
Z16110000491600; Youth Innovative Research Team of Capital Normal University and
Science Research Base Development Program of the Beijing Municipal Commission of
Education.}

\clearpage
\end{document}